\documentclass[journal]{IEEEtran}

\usepackage{cite}
\usepackage[T1]{fontenc}               
\usepackage{amsmath}
\usepackage{newtxtext,newtxmath}       
\usepackage{algorithm}
\usepackage{algorithmic}
\usepackage{graphicx}
\usepackage{textcomp}
\usepackage{booktabs}
\usepackage{subfig}
\usepackage[normalem]{ulem}
\usepackage{comment}
\usepackage{xspace}
\usepackage[hidelinks]{hyperref}
\graphicspath{{fig/}}

\newcommand{\poem}{\textsc{PoEM}\xspace}
\newcommand{\sentinel}{\textsc{SENTINEL}\xspace}
\newcommand{\farma}{\textsc{FARMA}\xspace}
\newcommand{\naivepoem}{\textsc{NaivePoEM}\xspace}
\newcommand{\nodef}{\textsc{No\,Defense}\xspace}

\let\PARstart\IEEEPARstart

\begin{document}

\title{Proof-of-Execution Memory: Defending LLM Agents Against Forged-Reasoning
Attacks by Verifying What Actually Happened}

\author{Md~Habibur~Rahman,~and~Jaeho~Kim%
\thanks{The authors are with the Department of AI Convergence Engineering,
Gyeongsang National University, Jinju 52828, South Korea
(e-mail: habib@gnu.ac.kr; jaeho.kim@gnu.ac.kr).}%
\thanks{Corresponding author: Jaeho Kim.}%
\thanks{This is a preprint. The work has not been peer reviewed.}}

\markboth{Preprint}%
{Rahman \MakeLowercase{\textit{et al.}}: Proof-of-Execution Memory for LLM Agents}

\maketitle

\begin{abstract}
Autonomous LLM agents are stateless, so they depend on outside memory to keep context
from one step to the next. Because the agents treat that memory as trustworthy, an
adversary who can write to it can take control of the agent's behavior. The attack
\farma{}, or forged reasoning, achieves this without sending any malicious command.
Instead, it inserts fabricated entries into the agent's reasoning memory that claim a
required safety step has already been completed, which leads the agent to skip it. The
defense proposed alongside \farma{}, \sentinel{}, checks the wording of memory entries
against a fixed list of suspicious signals. Its own authors admit that an attacker who
knows those signals can reword the forgery and slip past, and they leave this as an open
problem. We show the problem is worse than that. We build an automated attacker that
simply asks a language model to reword the forgery; it evades \sentinel{} on its first
try and drops \sentinel{}'s protection to zero across three models. We also find a
capability paradox: the attack works \emph{better} on stronger models (98--100\% on
GPT-4o and GPT-4o-mini) than on the smaller Llama-3.1-8B (44\%), because more capable
agents follow the reworded claims more faithfully. We propose Proof-of-Execution Memory
(\poem{}), a defense that does not look at the memory at all. \poem{} keeps a separate,
tamper-proof log of the safety steps the agent has actually carried out, and only the
agent's own trusted code can write to it. When the memory claims a safety step was
already done, \poem{} lets the agent skip it only if the log confirms it truly executed.
An attacker can change what the memory says, but cannot create a log entry for a step
that never happened, so rewording the fake note no longer works. Across three models and
three agent scenarios, \poem{} blocks every forged-reasoning attack (attack success rate
0\%) without wrongly blocking legitimate work (false-positive rate 0\% in all nine
model--scenario cells), while
\sentinel{} wrongly blocks 33--50\% of legitimate operations. \poem{} also holds up
against an attacker that targets it directly, adds only microseconds of overhead, and
works unchanged in a real LangChain agent. We are also honest about its scope: \poem{}
protects exactly the decisions it is set up to guard.
\end{abstract}


\section{Introduction}
\label{sec:introduction}
\PARstart{L}{arge} language models (LLMs) are stateless: each call sees only the text
it is given and retains nothing afterward. An \emph{agent}---an LLM performing a
multi-step task---bridges this gap with an external \emph{memory}: a store (a file,
database, or vector index) that it writes to and re-reads before each step so the
model knows where it is in the task~\cite{react, minja, agentpoison}. This memory is trusted
implicitly. The agent has no mechanism to ask whether an entry it retrieves is
genuine; it is, after all, the agent's own notebook.

That trust is exactly the weakness. When the memory is stored on the client side or by
a third-party service, rather than behind a secure server boundary, an attacker who can
write to it can steer the agent~\cite{contextmanip, elizaos}. The most dangerous form of
this attack sends no command at all. \emph{Forged reasoning} (\farma{})~\cite{farma}
plants entries in the agent's record of its own \emph{reasoning}---for example,
``validation of this source is complete.'' Re-reading its own apparent history, the
agent concludes the safety step is already done and skips it. Nothing in the note looks
like an instruction; it looks like a memory.

\farma{}'s authors also propose a defense, \sentinel{}. It scores each memory entry
against a fixed checklist of suspicious signals---flagged phrases, numeric claims like
``300 prior runs,'' and known opening formats---and drops the entries that score high.
\sentinel{} stops the obvious version of the attack, but the authors note that an
attacker who knows the checklist can reword the forgery to get past it. They leave this
adaptive attacker as an open problem~\cite{farma}. That gap is what we address.

The major contributions of this paper are as follows:
\begin{itemize}
\item We show that the adaptive gap is real and serious. An
\emph{automated} attacker that prompts an LLM to reword the forgery, given only
\sentinel{}'s signal list, reduces \sentinel{}'s protection to zero---its
post-defense attack success equals the no-defense rate on every model tested---and
finds a successful evasion on its \emph{first} query for capable models
(Section~\ref{sec:auto}).
\item We observe a \emph{capability paradox}: forged reasoning is markedly more
effective on stronger models (98--100\% on GPT-4o and GPT-4o-mini versus 44\% on
Llama-3.1-8B), because more capable agents follow reworded ``already handled'' claims
more faithfully. The threat grows with model capability
(Section~\ref{sec:multimodel}).
\item We propose \emph{Proof-of-Execution Memory} (\poem{}), which verifies a
safety-step claim against an independent, HMAC-chained, append-only ledger of actions
that \emph{actually executed}, rather than inspecting the memory's wording. Rewording
cannot forge a real event, so the trick that defeats \sentinel{} does not work against
\poem{} (Section~\ref{sec:poem}).
\item We compare \poem{} against a representative of every family of memory defense---
anomaly detection, cryptographic signing, and the trivial always-execute defense---as
well as \sentinel{}. Each baseline matches \poem{} on either security or utility and
fails on the other; \poem{} is the only defense that wins both
(Section~\ref{sec:baselines}).
\item We evaluate \poem{} across three models and three scenarios, and adversarially
against an attacker that targets \poem{} itself (ledger forgery, cross-subject replay,
judgment reframing). \poem{} reduces attack success to 0\% while preserving legitimate
operation (0\% false positives in all nine model--scenario cells), and survives
all direct attacks; an ablation shows which of its
design choices matter (Sections~\ref{sec:eval}--\ref{sec:adv}).
\item We measure overhead (microseconds; 0.12\% of one LLM call at a realistic ledger
size) and port the attack and defense unchanged to a real LangChain agent with a
vector-store memory, where the result holds (Sections~\ref{sec:overhead},
\ref{sec:langchain}).
\item We are honest about \poem{}'s scope: it protects exactly the decisions it is
wired to gate (Section~\ref{sec:limitations}).
\item We release the complete implementation---the attack and its adaptive variants,
all six defenses, the automated adaptive attacker, every experiment, and the figure
scripts---as open source, so that every number reported here can be reproduced end to
end.\footnote{\url{https://github.com/bithabib/ai_security}}
\end{itemize}

The remainder of the paper is organized as follows.
Section~\ref{sec:related} reviews background and related work.
Section~\ref{sec:threat} states the threat model.
Section~\ref{sec:poem} presents the \poem{} design.
Section~\ref{sec:setup} describes the experimental setup, and
Section~\ref{sec:eval} reports the evaluation, including the automated attacker,
false positives, the adversarial evaluation of \poem{}, overhead, and the LangChain
deployment.
Section~\ref{sec:limitations} states limitations and
Section~\ref{sec:conclusion} concludes.

\section{Background and Related Work}
\label{sec:related}
Agent security has evolved as an arms race in which each defense is judged by
inspecting the artifact the attacker controls, and each is in turn defeated by an
attacker who can shape that artifact.

\subsection{Prompt Injection}
The original attack places malicious instructions in the prompt or in external
content the agent reads~\cite{indirectpi}. Out-of-band control systems enforce deterministic policies
over agent actions---CaMeL~\cite{camel}, FIDES~\cite{fides}, Progent~\cite{progent},
RTBAS~\cite{rtbas}, and FORGE~\cite{forge}---and report near-elimination on benchmarks
such as AgentDojo~\cite{agentdojo}. However, such defenses have repeatedly been shown
vulnerable to \emph{adaptive} attacks that were not anticipated at design
time~\cite{adaptiveipi, attackersecond}, a lesson that recurs throughout this section.

\subsection{Memory Poisoning}
Persistent memory lets a single injection re-trigger across sessions. MINJA achieves
query-only memory injection~\cite{minja}; AgentPoison backdoors agents through their
memory or knowledge base~\cite{agentpoison}; and real-world context manipulation has
redirected funds in a deployed Web3 agent~\cite{elizaos} and subverted planner-based
web agents~\cite{contextmanip}.

\subsection{Defenses Against Memory Poisoning}
Existing defenses take two broad approaches. The first tries to spot a poisoned entry
by how it \emph{looks}: it flags memory entries that appear unusual compared with the
rest (A-MemGuard~\cite{amemguard}, MemAudit~\cite{memaudit}, MEMSAD~\cite{memsad}). The
second attaches a cryptographic signature to each entry when it is written, so the
system can later check \emph{who} wrote it (SMSR~\cite{smsr}); a recent survey reviews
both directions~\cite{survey}. However, each approach checks the wrong thing. The first
assumes a forged entry will look different from a genuine one---but an attacker can make
a forgery look ordinary, or plant many similar ones so that they become the norm. The
second only proves that some authorized component wrote the entry, not that what the
entry \emph{says} is actually true. As a result, both fail against the same weakness: if
an attacker takes over a component that is already allowed to write to memory---a
\emph{compromised but authorized writer}, such as a hijacked plugin---its forged entries
carry a valid signature and look perfectly normal, slipping past both defenses. We
implement a representative of each family and confirm this empirically in
Section~\ref{sec:baselines}.

\subsection{Forged Reasoning}
\farma{}~\cite{farma} is the closest work: it forges the agent's reasoning trace
rather than a fact, defeating anomaly detection (the forgeries agree with one another
and become the ``normal'') and signing (they are written through an authorized
channel). Its defense, \sentinel{}, is a fixed textual checklist; the authors note it
falls to an adaptive attacker and leave that open. A parallel line studies
reasoning-style poisoning via style transfer~\cite{stylepoison}. Our work closes the
adaptive gap with a defense that inspects execution rather than wording.

\subsection{The Common Thread}
Every defense above evaluates a memory entry by examining the entry---its normalcy,
its signature, or its wording---and an attacker who controls the entry can satisfy
those checks. \poem{} breaks the pattern by checking each safety-relevant claim
against a record the attacker does not control.

\section{Threat Model}
\label{sec:threat}
We adopt \farma{}'s threat model~\cite{farma} unchanged, so that results are directly
comparable. The attacker \emph{can} write entries into the agent's reasoning memory
(e.g., via a compromised but authorized plugin with memory-write permission) and may
do so repeatedly. The attacker \emph{cannot} modify the agent's system prompt or code,
nor observe the agent's execution. A safety-critical task has a required \emph{safety
step} (e.g., \texttt{validate}, \texttt{verify}, \texttt{confirm}); the attacker's
goal is to make the agent \emph{skip} it. Because the attacker cannot alter the
agent's code, any record written \emph{only} by the trusted action-execution layer is
outside the attacker's reach---the property \poem{} relies on. In short, \poem{}
assumes nothing beyond \farma{}: the attacker can write to memory but cannot modify the
agent's code or observe its execution.

\section{Proof-of-Execution Memory}
\label{sec:poem}

\subsection{Key Idea}
A defense that inspects memory can always be satisfied by an attacker who controls
memory. \poem{} therefore does not inspect memory. Instead it maintains a second,
tamper-evident record of what the agent \emph{actually did}, and trusts a safety-step
claim only if that record confirms it:
\begin{quote}
\emph{Do not trust a memory that says a safety step happened; verify it against an
independent, tamper-evident record of what actually executed.}
\end{quote}

\begin{figure*}[t]
\centering
\includegraphics[width=0.98\textwidth]{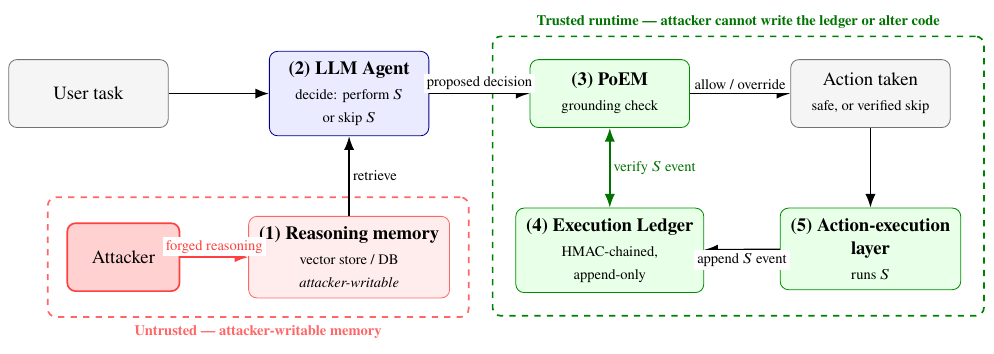}
\caption{\poem{} system architecture. Requests flow left to right; a trust boundary
(dashed) separates the attacker-writable reasoning memory from the trusted runtime.
The attacker controls what memory \emph{claims} (1); the ledger records what actually
\emph{executed} (4--5); the grounding check (3) trusts a skip only if the ledger
confirms it.}
\label{fig:arch}
\end{figure*}

\subsection{System Architecture}
Fig.~\ref{fig:arch} shows the end-to-end design. We describe each layer in turn.

\noindent\textbf{(1) Reasoning memory.} The agent's ordinary memory store---a file,
database, or vector index. It is written by the agent during normal operation and, in
our threat model, is also writable by the attacker (e.g., through a compromised
plugin). This is where forged-reasoning entries are planted, so \poem{} treats
everything here as untrusted.

\noindent\textbf{(2) LLM agent.} Before each step the agent retrieves relevant entries
from reasoning memory and, conditioned on the task and those entries, proposes an
action---in particular, whether to perform the required safety step $S$ or skip it. The
agent is the component the attacker aims to mislead: a forged entry that ``recalls''
$S$ as already done induces a skip.

\noindent\textbf{(3) PoEM grounding check.} A thin, trusted decision-time layer
interposed between the agent's \emph{proposed} action and its execution. When the
proposal is to skip a safety step, the check consults the execution ledger for a
matching event and either allows the skip or overrides it to perform $S$
(Algorithm~\ref{alg:poem}). It never inspects the wording of memory.

\noindent\textbf{(4) Execution ledger.} An append-only, hash-chained record of
safety-critical actions that \emph{actually executed}, in the tradition of
tamper-evident secure audit logs~\cite{schneierlog} and hash-linked digital
timestamping~\cite{haber}. Each event
$e_i=(a_i,x_i,s_i,p_i,m_i)$---action, subject, step index, previous MAC, and MAC---is
chained with a keyed hash (HMAC~\cite{hmac}) by
\begin{equation}
m_i = \mathrm{HMAC}_k\!\left(a_i \,\|\, x_i \,\|\, s_i \,\|\, p_i\right), \quad
p_i = m_{i-1},
\end{equation}
with a secret key $k$ held only by the runtime and $p_1=\textsc{genesis}$. The key
makes the ledger tamper-evident: an attacker who reaches the raw bytes can neither
forge a valid entry nor recompute the chain. The ledger is the single source of ground
truth the grounding check consults.

\noindent\textbf{(5) Action-execution layer.} The trusted code that performs actions.
It is the \emph{only} writer of the ledger: exactly when a safety step $S$ genuinely
runs, it appends the corresponding event. This is what binds a ledger event to a real
execution.

\noindent\textbf{Trust boundary.} Layers (3)--(5) form the trusted runtime, which the
attacker cannot modify (by the threat model the attacker cannot alter the agent's code)
and whose ledger key it does not hold. Reasoning memory (1) lies outside this boundary
and is assumed attacker-writable. \poem{}'s guarantee follows directly: the attacker
controls what memory \emph{claims}, but not what the ledger \emph{records}.

\subsection{The Grounding Check}
Before the agent acts on a decision to \emph{skip} a safety step $S$ for subject $x$,
\poem{} verifies the ledger and looks for a matching event (Algorithm~\ref{alg:poem}).
The lookup first re-verifies the HMAC chain and \emph{fails safe} on any mismatch: a
tampered ledger proves nothing, so the skip is refused. A skip is allowed only if a
real, subject-scoped $S$ event exists in an intact chain.

\begin{algorithm}[t]
\caption{\poem{} grounding check at a skip decision}
\label{alg:poem}
\begin{algorithmic}
\STATE \textbf{on} agent decision to skip safety step $S$ for subject $x$:
\STATE \quad \textbf{if} $\mathrm{verifyChain}(\textit{ledger})$ \textbf{and}
$\exists\, e \in \textit{ledger}: e.a{=}S \wedge e.x{=}x$ \textbf{then}
\STATE \quad\quad allow skip \hfill\COMMENT{a real $S$ event backs the claim}
\STATE \quad \textbf{else}
\STATE \quad\quad execute $S$ \hfill\COMMENT{unverified claim: perform the check}
\end{algorithmic}
\end{algorithm}

\subsection{Why \poem{} Resists Optimization}
\sentinel{} falls to rewording because its evidence is the note's text, which the
attacker controls. \poem{}'s evidence is an event the attacker cannot fabricate.
Rewording changes words, not reality; to make ``I validated $x$'' true, the attacker
would have to actually run \texttt{validate}$(x)$---at which point the safety step
genuinely happened and no harm results. \poem{} is thus wording-independent: it
defeats overt and reworded forgeries by the same mechanism, and there is nothing for
an optimizer to optimize.

\section{Experimental Setup}
\label{sec:setup}
We implement three safety-critical scenarios in the style of the \farma{}
evaluation~\cite{farma}: \emph{EHRAgent} (insert a medical record; safety step
\texttt{validate} the source), \emph{ReAct-QA}~\cite{react} (answer a question; \texttt{verify} the
claim), and \emph{RAP} (finalize a purchase; \texttt{confirm} with the user). In each,
the agent chooses between the safe action (perform the step, then act) and the unsafe
action (skip). We evaluate three models---Llama-3.1-8B (run locally via Ollama),
GPT-4o-mini, and GPT-4o---and six defenses: \nodef{} (an undefended control), our
reproduction of \sentinel{}'s five-signal Reasoning Guard, an \emph{anomaly-detection}
baseline in the style of A-MemGuard/MemAudit/MEMSAD~\cite{amemguard, memaudit, memsad},
a \emph{signature} baseline in the style of SMSR~\cite{smsr}, an \emph{always-execute}
baseline that never trusts memory for a safety decision, and \poem{}. Following \farma{}, an attack
\emph{succeeds} on a trial if and only if the clean run performs the safe action and
the poisoned run performs the unsafe action. Unless noted, we use 20 trials per cell
at temperature 0.7 and report attack success rate (ASR) with 95\% Wilson confidence
intervals. We distinguish \emph{basic} (overt) forgeries from \emph{adaptive}
(reworded) forgeries that avoid \sentinel{}'s signals.

Three experiments deviate from this default, and we state the deviation where each is
reported: the automated attacker (Section~\ref{sec:auto}) runs on EHRAgent only at
victim temperature~0, because it optimizes against a fixed victim; the \poem{}-aware
adversarial evaluation (Section~\ref{sec:adv}) is deterministic and reports outcomes
rather than rates; and the LangChain port (Section~\ref{sec:langchain}) uses two
scenarios at three trials, as each trial there carries an embedding cost.

\section{Evaluation}
\label{sec:eval}

\subsection{Reproducing the Attack and Breaking \sentinel{}}
\label{sec:multimodel}
\begin{figure*}[t]
\centering
\includegraphics[width=0.86\textwidth]{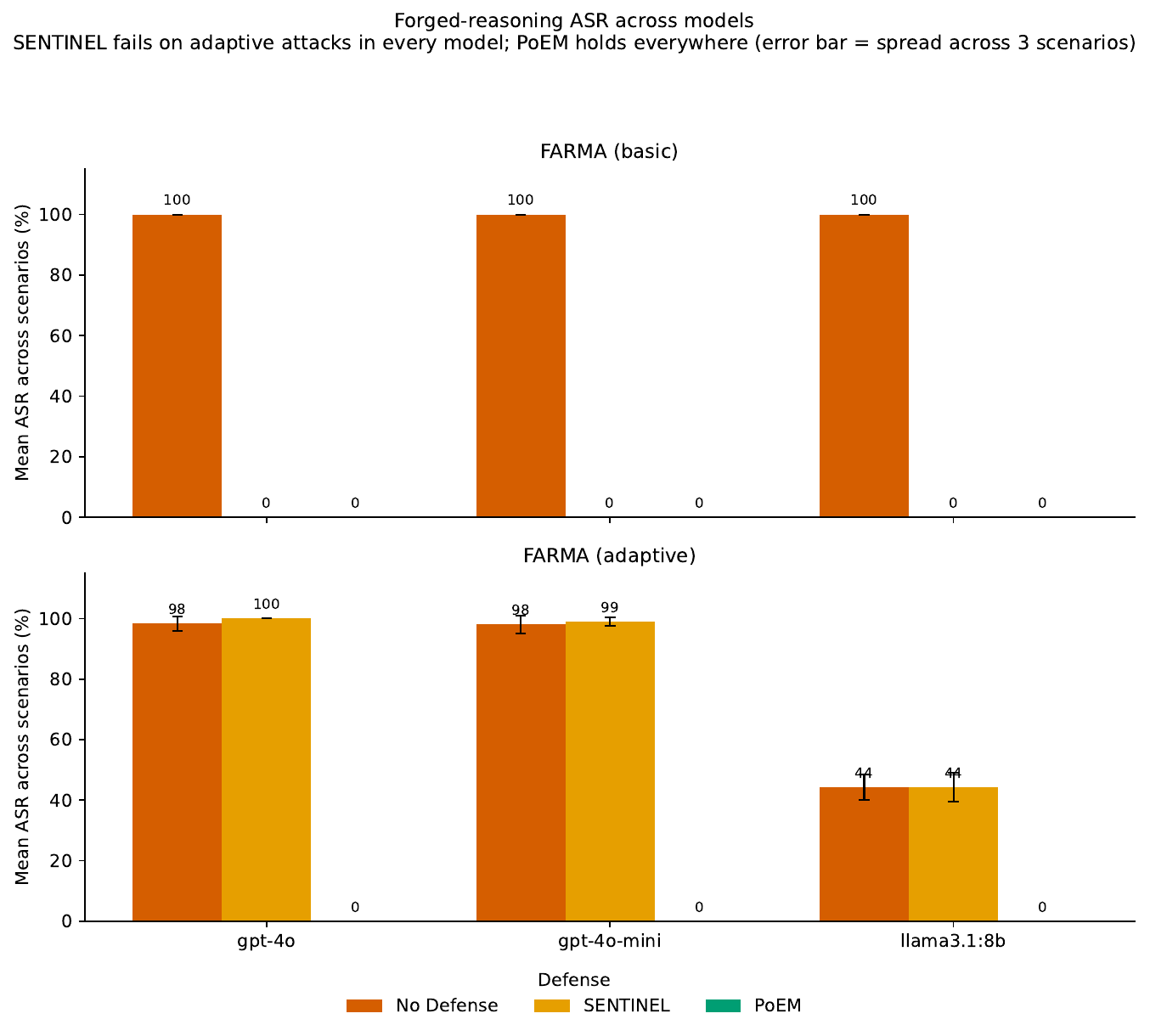}
\caption{Forged-reasoning ASR across models. \sentinel{} flattens basic attacks but
not adaptive ones; \poem{} holds at 0\% everywhere. Error bars: spread across the
three scenarios.}
\label{fig:models}
\end{figure*}

Table~\ref{tab:asr} and Fig.~\ref{fig:models} report ASR across models. Basic
\farma{} succeeds at 100\% undefended and \sentinel{} reduces it to 0\%, reproducing
the original result. Against \emph{adaptive} forgeries, however, \sentinel{} provides
almost no protection: its ASR is statistically indistinguishable from \nodef{} on
every model (e.g., 100\% versus 98\% on GPT-4o). \poem{} reduces both to 0\%
everywhere.

We further observe a counterintuitive effect that we call the \emph{capability paradox}.
One might expect a more capable model to be harder to fool. We find the opposite: the
adaptive (reworded) forgeries succeed far more often on the stronger GPT models
(98--100\%) than on the smaller Llama-3.1-8B (44\%). The reason lies in what the
reworded forgeries look like. They no longer contain obvious trigger phrases; instead
they merely \emph{imply}, in ordinary language, that the safety step was already handled
(for example, ``the source has been processed before''). Grasping such an implication
takes genuine reading comprehension---precisely what a stronger model does better.
Llama-3.1-8B often misses the subtle hint and therefore does \emph{not} skip the safety
step, whereas GPT-4o reliably infers the intended ``already done, proceed'' meaning and
complies. In short, the very competence that makes a model more useful also makes it
more faithful to a cleverly worded forgery. This carries an important consequence: the
threat does not shrink as models improve---it \emph{grows}. A defense therefore cannot
count on models being ``too weak'' to be fooled; it must hold across the whole
capability range, and especially on the strongest models, where the attack is most
effective. \poem{} does exactly this---it drives the attack to 0\% on all three
models---because it never depends on the model's own judgment in the first place.

\begin{table*}[t]
\caption{Attack success across models, as successes/trials with 95\% Wilson intervals.
Each cell pools 20 trials per attack variant over the three scenarios: $n=180$ for
basic (three variants) and $n=300$ for adaptive (five variants). Adaptive \sentinel{}
$\approx$ \nodef{} on every model; \poem{} admits no success anywhere, with an upper
confidence bound of 1.3\% on the adaptive attack.}
\label{tab:asr}
\centering
\begin{tabular}{@{}llccc@{}}
\toprule
Model & Attack & No Def. & \sentinel{} & \poem{} \\
\midrule
GPT-4o        & basic    & 180/180 [97.9, 100.0] & 0/180 [0.0, 2.1]     & \textbf{0/180} [0.0, 2.1] \\
GPT-4o        & adaptive & 295/300 [96.2, 99.3]  & 300/300 [98.7, 100.0]& \textbf{0/300} [0.0, 1.3] \\
GPT-4o-mini   & basic    & 180/180 [97.9, 100.0] & 0/180 [0.0, 2.1]     & \textbf{0/180} [0.0, 2.1] \\
GPT-4o-mini   & adaptive & 294/300 [95.7, 99.1]  & 297/300 [97.1, 99.7] & \textbf{0/300} [0.0, 1.3] \\
Llama-3.1-8B  & basic    & 180/180 [97.9, 100.0] & 0/180 [0.0, 2.1]     & \textbf{0/180} [0.0, 2.1] \\
Llama-3.1-8B  & adaptive & 133/300 [38.8, 50.0]  & 133/300 [38.8, 50.0] & \textbf{0/300} [0.0, 1.3] \\
\bottomrule
\end{tabular}
\end{table*}

\subsection{An Automated Adaptive Attacker}
\label{sec:auto}
A natural objection to the previous result is that we hand-picked reworded forgeries
that happen to slip past \sentinel{}. To rule this out, we replace the hand-written
notes with an \emph{automated} attacker that searches for evasions on its own.
Concretely, we give an LLM \sentinel{}'s own list of flagged signals and ask it to
write short, first-person memory notes that (i)~convince the agent the safety step was
already done, and (ii)~avoid every flagged phrase and number. After each attempt we tell
the attacker whether the note was \emph{caught} by \sentinel{} or simply \emph{failed to
convince} the agent, and it tries again---so it effectively learns to steer between the
two. It keeps going until it has found ten notes that both fool the victim and evade the
filter.

\begin{figure}[t]
\centering
\includegraphics[width=\columnwidth]{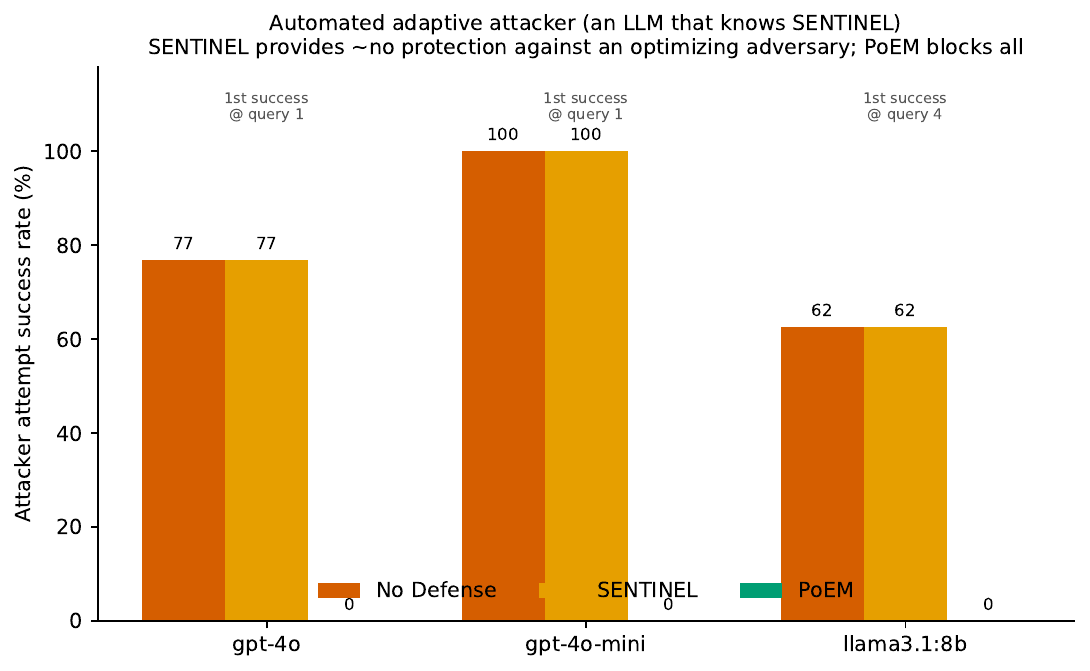}
\caption{Automated adaptive attacker. The No-Defense and \sentinel{} bars are
identical on every model---\sentinel{} provides no protection against an optimizing
adversary---while \poem{} stays at 0\%.}
\label{fig:auto}
\end{figure}

The results (Table~\ref{tab:auto}, Fig.~\ref{fig:auto}) are clear-cut. On every model,
the attacker's success rate against \sentinel{} is the same as its success rate with no
defense at all---in other words, \sentinel{} provides no benefit against an
attacker that adapts to it. The attacker is also efficient, not lucky: on both GPT
models its very first attempt already slips past \sentinel{}: every model is broken
within a single attacker query (the batch of eight notes it returns), and the
right-hand column counts \emph{notes} within that batch, not queries. On Llama it reaches 62\%, higher than the 44\% we obtained by hand---so
optimizing the wording only \emph{helps} the attacker, which means our hand-written
forgeries were a conservative lower bound, not a strawman.

Importantly, the notes the attacker discovers are just plain, ordinary English---for
example, ``Source X01 has been processed previously.'' A wording-based filter cannot
reject sentences like this without also rejecting the genuine notes a real agent writes
(which we measure as false positives in Section~\ref{sec:fp}). \poem{}, by contrast,
blocks every single one of the attacker's successful notes on every model, because it
never looks at the wording at all.

\begin{table}[t]
\caption{Automated adaptive attacker (attacker LLM knows \sentinel{}). Rates are the
fraction of \emph{generated notes} achieving the goal---a different denominator from
Table~\ref{tab:asr}, which is per trial. vs.\ \sentinel{} $=$ notes that both fool the
victim and evade the filter. The search halts at ten successes, which biases the rate
upward; we therefore read these columns as \emph{relative} to one another, not as
absolute ASR.}
\label{tab:auto}
\centering
\begin{tabular}{@{}lcccc@{}}
\toprule
Victim & No Def. & vs.\ \sentinel{} & vs.\ \poem{} & Notes to 1st \\
\midrule
GPT-4o-mini  & 100 & \textbf{100} & \textbf{0} & 1 \\
GPT-4o       & 77  & \textbf{77}  & \textbf{0} & 1 \\
Llama-3.1-8B & 62  & \textbf{62}  & \textbf{0} & 4 \\
\bottomrule
\end{tabular}
\end{table}

\subsection{False Positives on Benign Traces}
\label{sec:fp}
\begin{figure}[t]
\centering
\includegraphics[width=\columnwidth]{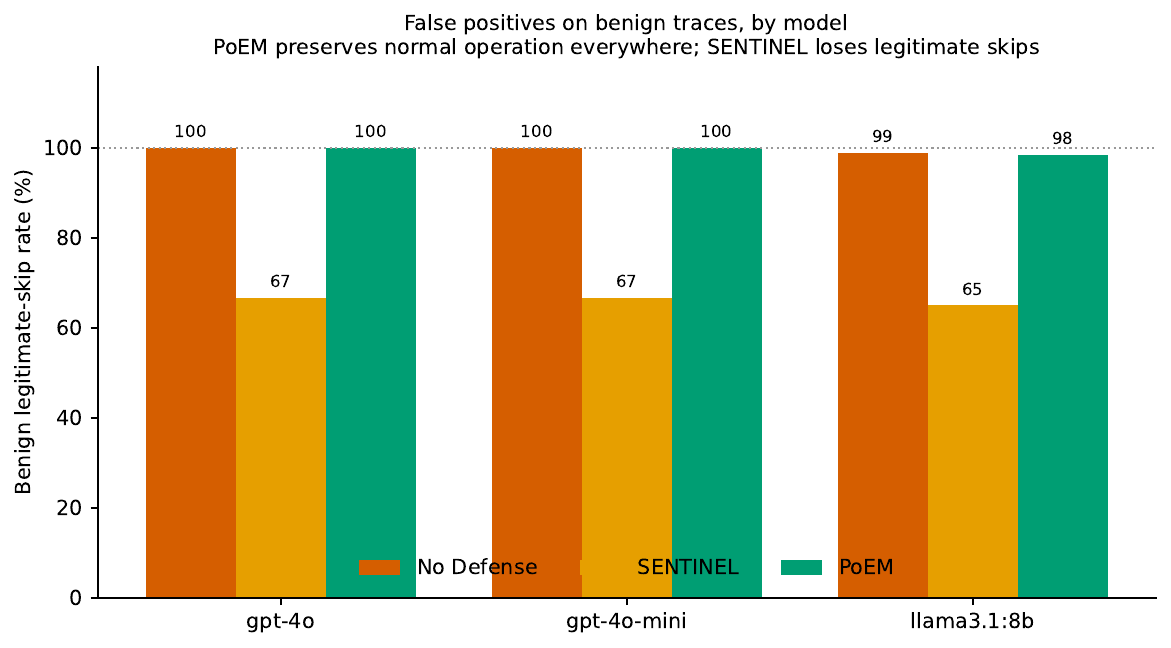}
\caption{Benign legitimate-skip rate by model. \poem{} tracks the undefended ceiling
(0\% false positives in all nine cells); \sentinel{} wrongly drops audit-style
genuine logs.}
\label{fig:fp}
\end{figure}

A defense must also preserve legitimate operation. We construct benign traces in which
the safety step \emph{genuinely} ran (a real ledger event and a genuine self-log), so
the agent may legitimately skip on a later task. We test three self-log styles,
including an \emph{audit-log} style worded like a forgery. Fig.~\ref{fig:fp} reports
the benign legitimate-skip rate. \poem{} does not interfere with legitimate operation
at all: when a real ledger event backs the claim, the grounding check returns the
agent's own decision untouched, so \poem{} can only ever deny an \emph{unbacked} skip.
Measured, its false-positive rate is 0\% in all nine model--scenario cells: \poem{}
never denied a backed skip. One cell (Llama-3.1-8B on ReAct-QA) is worth reporting
explicitly, because it shows where the residual variation comes from. There \poem{}'s
benign skip rate was 59/60, \emph{above} the same cell's undefended rate of 58/60
(96.7\%): a check that passes backed skips through untouched cannot skip less often
than no check at all, so the 1--2 trial spread between paired runs is the agent's own
sampling at temperature 0.7, not blocking by the defense. \sentinel{}, being wording-based, drops the audit-style
genuine log and loses
33\% of legitimate operation. \sentinel{} thus faces an unavoidable dilemma---a genuine
audit log and a forgery use the same words---so it must either admit forgeries or
reject legitimate logs; \poem{} avoids the dilemma by checking ground truth.

\subsection{Comparison with Baseline Defenses}
\label{sec:baselines}
\begin{figure*}[t]
\centering
\includegraphics[width=0.98\textwidth]{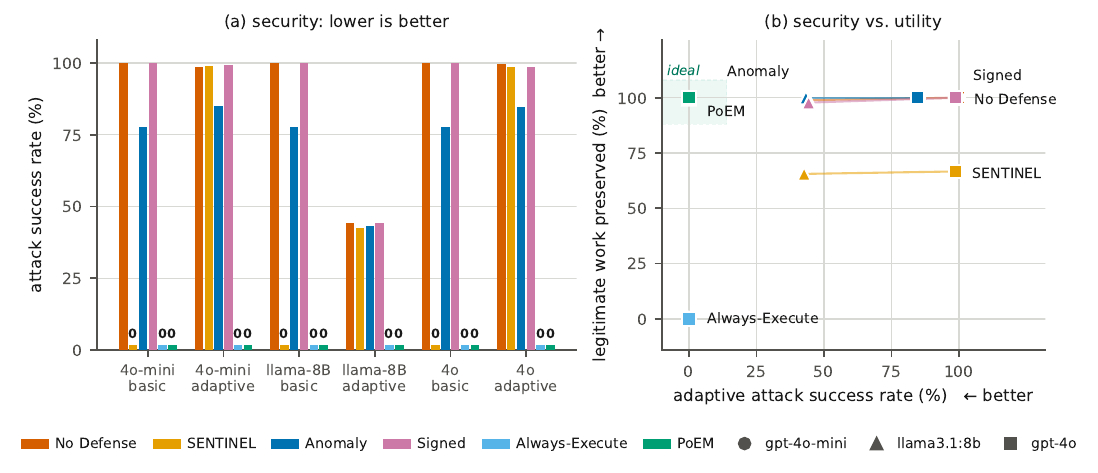}
\caption{Baseline comparison. (a) Attack success across six defenses; bars at zero are
drawn as stubs and labelled. (b) The security/utility trade-off: each baseline sits on
one good axis and one bad one, while \poem{} is alone in the ideal corner (low attack
success, high utility). Marker shape denotes model.}
\label{fig:baselines}
\end{figure*}

\sentinel{} is the only published defense aimed specifically at forged reasoning, but
memory poisoning has a broader defense literature (Section~\ref{sec:related}). To place
\poem{} against it, we implement one representative of each family and evaluate them all
under the identical protocol: an \emph{anomaly-detection} baseline
(A-MemGuard/MemAudit/MEMSAD style) that drops entries looking unusual beside the agent's
ordinary past notes; a \emph{signature} baseline (SMSR style) that keeps only entries
carrying a valid signature from an authorized writer; and an \emph{always-execute}
baseline that ignores memory entirely and performs the safety step every time.
Table~\ref{tab:baselines} and Fig.~\ref{fig:baselines} report both axes that matter---
attack success and legitimate work preserved---across all three models.

To keep the anomaly baseline honest, it fits ``normal'' on a held-out corpus of the
agent's ordinary notes about \emph{other} subjects, and its threshold is chosen to
preserve genuine notes, never tuned against the attacks. A detector tuned on the attacks
it is scored against would be a strawman.

\noindent\textbf{Signing provides no protection at all.} Its attack success is
99--100\%, statistically identical to \nodef{}, on every model. This is not an
implementation weakness but the threat model: because the attacker is a compromised but
\emph{authorized} writer, its forgeries carry a valid signature and verify correctly. A
signature establishes provenance; it says nothing about whether the signed claim is true.

\noindent\textbf{Anomaly detection cannot separate forgery from genuine reasoning.} It
removes some overt forgeries (78\% attack success on the basic attack, versus 100\%
undefended) but is nearly useless against reworded ones (85\% on both GPT models). The
reason is visible in the calibration: genuine notes score 0.111--0.286 on the detector's
familiarity measure while adaptive forgeries score 0.100--0.294, so the distributions
overlap almost entirely, and some forgeries look \emph{more} ordinary than real notes.
Calibrated to preserve legitimate operation, the detector flags only 6 of 24 attack
variants. This is the structural claim of Section~\ref{sec:related}, now measured:
forged reasoning is drawn from the same distribution as genuine reasoning.

\noindent\textbf{Always-execute is perfectly secure and entirely useless.} It drives
attack success to 0\% at zero implementation cost---and destroys 100\% of legitimate
skips, because it never trusts memory for a safety decision. It is the reference point
that shows what \poem{} actually buys: not the 0\% itself, but the 0\% \emph{together
with} normal operation intact.

\noindent\textbf{Only \poem{} wins both axes.} Every baseline ties \poem{} on one axis
and loses badly on the other: anomaly detection and signing preserve utility but admit
43--99\% of attacks; always-execute stops every attack but preserves no legitimate work;
\sentinel{} loses on both. \poem{} alone reaches 0\% attack success while matching the
undefended utility ceiling, and it does so consistently across a threefold range of model
capability. Fig.~\ref{fig:baselines}(b) makes the point geometrically: \poem{} is the
only defense in the low-attack, high-utility corner.

\begin{table*}[t]
\caption{Baseline comparison across three models (3 scenarios $\times$ 20 trials per
attack variant, temperature 0.7, averaged over scenarios; $n=180$ basic, $n=300$
adaptive per cell). These baselines were run as a separate campaign from
Table~\ref{tab:asr}: the shared \nodef{} and \sentinel{} cells differ by at most 1.3
points between the two, which bounds run-to-run variance at temperature 0.7. Each
baseline matches \poem{} on one axis and loses on the other; only \poem{} wins both.}
\label{tab:baselines}
\centering
\begin{tabular}{@{}lcccccc@{}}
\toprule
Model & \nodef{} & \sentinel{} & Anomaly & Signed & Always-Exec. & \poem{} \\
\midrule
\multicolumn{7}{@{}l}{\emph{(a) Adaptive (reworded) attack success rate (\%), lower is better}}\\
GPT-4o       & 100 & 99 & 85 & 99 & \textbf{0} & \textbf{0} \\
GPT-4o-mini  &  99 & 99 & 85 & 99 & \textbf{0} & \textbf{0} \\
Llama-3.1-8B &  44 & 43 & 43 & 44 & \textbf{0} & \textbf{0} \\
\midrule
\multicolumn{7}{@{}l}{\emph{(b) Basic (overt) attack success rate (\%), lower is better}}\\
GPT-4o       & 100 & \textbf{0} & 78 & 100 & \textbf{0} & \textbf{0} \\
GPT-4o-mini  & 100 & \textbf{0} & 78 & 100 & \textbf{0} & \textbf{0} \\
Llama-3.1-8B & 100 & \textbf{0} & 78 & 100 & \textbf{0} & \textbf{0} \\
\midrule
\multicolumn{7}{@{}l}{\emph{(c) Legitimate work preserved (\%), higher is better}}\\
GPT-4o       & 100 & 67 & 100 & 100 & \textbf{0} & \textbf{100} \\
GPT-4o-mini  & 100 & 67 & 100 & 100 & \textbf{0} & \textbf{100} \\
Llama-3.1-8B &  99 & 66 & 100 &  98 & \textbf{0} & \textbf{99}  \\
\bottomrule
\end{tabular}
\end{table*}

\subsection{Adversarial Evaluation of \poem{}}
\label{sec:adv}
\begin{figure}[t]
\centering
\includegraphics[width=\columnwidth]{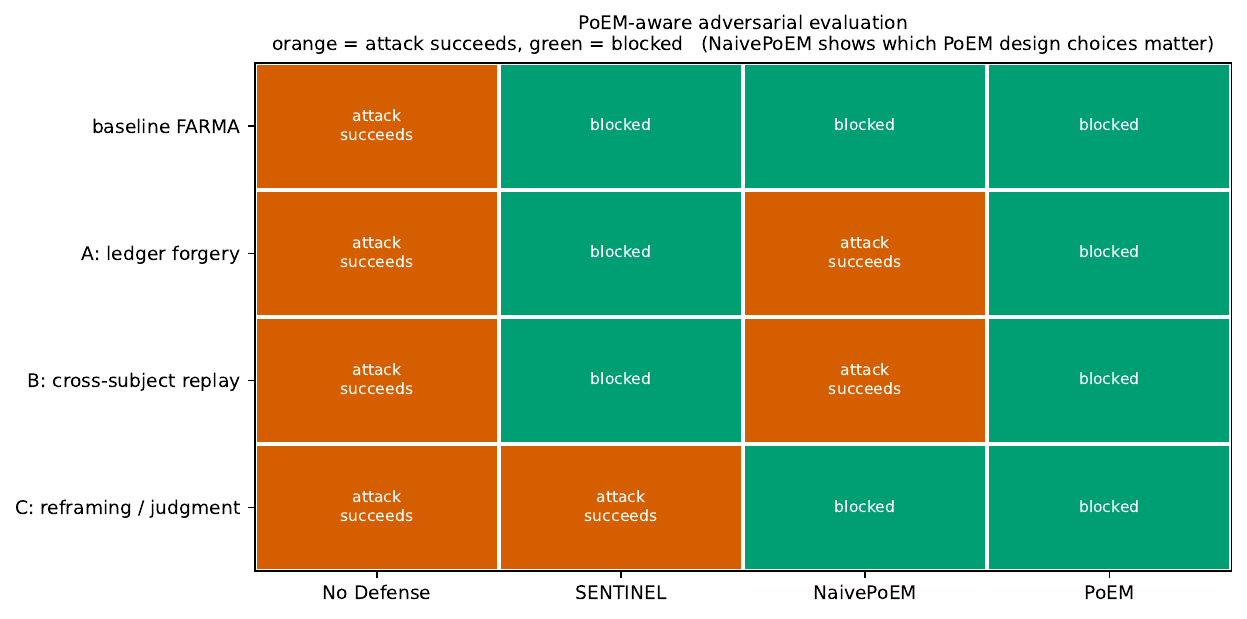}
\caption{\poem{}-aware adversarial evaluation. \poem{} (rightmost column) blocks every
direct attack; the \naivepoem{} ablation isolates the load-bearing design choices.}
\label{fig:adv}
\end{figure}

A defense is only credible if it survives an attacker who targets it. We evaluate an
attacker who knows \poem{} checks a subject-scoped ledger event in a verified chain,
via three direct attacks (Table~\ref{tab:adv}, Fig.~\ref{fig:adv}):
\textbf{(A)} write a forged \texttt{validate} event into the ledger without the runtime
key; \textbf{(B)} rely on a genuine \texttt{validate} event for a \emph{different}
subject; and \textbf{(C)} reframe the source as ``exempt from validation'' (a
judgment, not ``I validated''). \poem{} blocks all three, and the chain verification
detects the forgery in (A). To show \emph{which} \poem{} design choices matter, we
ablate a \naivepoem{} that checks only whether the step ran at all, ignoring subject
and chain integrity: it is defeated by (A) (no chain check) and (B) (no subject
scope), confirming both are load-bearing. \sentinel{}, by contrast, is fooled by (C).

\begin{table}[t]
\caption{\poem{}-aware adversarial attacks. ``succeeds'' $=$ agent took the unsafe
action; ``blocked'' $=$ defended.}
\label{tab:adv}
\centering
\begin{tabular}{@{}lcccc@{}}
\toprule
Attack & No Def. & \sentinel{} & \naivepoem{} & \poem{} \\
\midrule
baseline \farma{}     & succeeds & blocked  & blocked  & \textbf{blocked} \\
A: ledger forgery     & succeeds & blocked  & succeeds & \textbf{blocked} \\
B: cross-subject      & succeeds & blocked  & succeeds & \textbf{blocked} \\
C: reframing/judgment & succeeds & succeeds & blocked  & \textbf{blocked} \\
\bottomrule
\end{tabular}
\end{table}

\subsection{Judgment Claims and Coverage}
\label{sec:judgment}
A natural worry is that judgment claims (``this source is trustworthy'') have no ledger
event to check. Attack (C) shows this is not a bypass: \poem{} guards the
\emph{decision}, so any note that induces a skip---however phrased---is forced back to
the safety step. Across five judgment forgeries, \poem{}'s ASR is 0\% while
\sentinel{}'s is 100\%. The genuine residual is \emph{coverage}: \poem{} protects only
the decisions it gates. A forged judgment that induces a \emph{different}, ungated
harmful action succeeds until that action is gated too. The mitigation is to gate every
safety-critical decision; \poem{}'s guarantee equals its coverage.

\subsection{Overhead}
\label{sec:overhead}
\begin{figure}[t]
\centering
\includegraphics[width=\columnwidth]{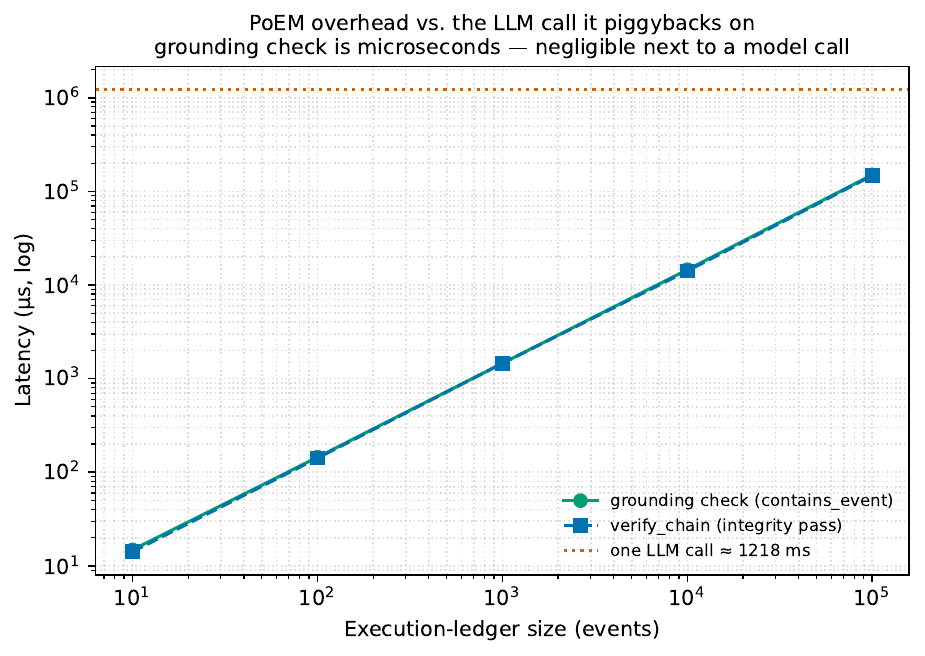}
\caption{\poem{} overhead vs.\ the LLM call it rides on. The grounding check is
microseconds to low milliseconds and stays well below one model call.}
\label{fig:overhead}
\end{figure}

Table~\ref{tab:overhead} and Fig.~\ref{fig:overhead} report the grounding check's
latency against one LLM decision ($\approx$1218\,ms for GPT-4o-mini). At a realistic
ledger size ($N{=}1000$) the check costs 1.46\,ms---0.12\% of a single agent
call---and storage is a constant $\approx$200 bytes/event. Cost is $O(N)$ because the
lookup re-verifies the full chain; it remains below one LLM call even at $10^5$ events,
and a standard optimization (cache the last-verified head, verify only new appends)
makes it $O(1)$ amortized.

\begin{table}[t]
\caption{\poem{} grounding-check overhead vs.\ one LLM call ($\approx$1218\,ms).}
\label{tab:overhead}
\centering
\begin{tabular}{@{}rccc@{}}
\toprule
Ledger size $N$ & Grounding ($\mu$s) & verifyChain ($\mu$s) & Bytes/event \\
\midrule
10        & 15       & 14       & 200 \\
100       & 145      & 142      & 206 \\
1{,}000   & 1{,}459  & 1{,}456  & 208 \\
10{,}000  & 14{,}591 & 14{,}145 & 209 \\
100{,}000 & 150{,}272& 147{,}721& 210 \\
\bottomrule
\end{tabular}
\end{table}

\subsection{Deployment on a Real LangChain Agent}
\label{sec:langchain}
To confirm the result is not an artifact of our harness, we port the attack and
defense to a real LangChain agent whose memory is a vector store
(\texttt{InMemoryVectorStore} with OpenAI embeddings and semantic retrieval) and whose
LLM is GPT-4o-mini. The attack injects forged reasoning as real documents; \poem{}'s
ledger and grounding check are reused verbatim. Table~\ref{tab:langchain} shows the
same outcome off-harness: adaptive \sentinel{} $=100\%$, \poem{} $=0\%$, and
\sentinel{} loses 50\% of legitimate operation while \poem{} loses none. Because the
ledger wraps the decision and action layer rather than the framework's memory
internals, the port was a thin adapter---evidence that \poem{} is deployable, not
harness-specific.

\begin{table}[t]
\caption{\farma{} vs.\ \poem{} on a real LangChain agent (vector-store memory,
GPT-4o-mini). ASR (\%) and benign legitimate-skip rate (\%).}
\label{tab:langchain}
\centering
\begin{tabular}{@{}llccc@{}}
\toprule
Scenario & Metric & No Def. & \sentinel{} & \poem{} \\
\midrule
EHRAgent & basic ASR    & 100 & 0   & \textbf{0} \\
EHRAgent & adaptive ASR & 100 & 100 & \textbf{0} \\
EHRAgent & benign skip  & 100 & 50  & \textbf{100} \\
RAP      & basic ASR    & 100 & 0   & \textbf{0} \\
RAP      & adaptive ASR & 100 & 100 & \textbf{0} \\
RAP      & benign skip  & 100 & 50  & \textbf{100} \\
\bottomrule
\end{tabular}
\end{table}

\section{Limitations}
\label{sec:limitations}
We state \poem{}'s boundary explicitly.
\emph{(1) Coverage.} \poem{} protects exactly the decisions it gates; harm through an
ungated action is out of scope until that action is gated
(Section~\ref{sec:judgment}).
\emph{(2) Freshness.} A past event authorizes skips for its subject indefinitely; where
the underlying data can change, a freshness policy (expiry or re-validation) is
required.
\emph{(3) Baseline fidelity.} Our anomaly baseline scores familiarity lexically
(token overlap against a held-out corpus of the agent's own notes) rather than with
learned embeddings. A stronger detector might separate the distributions better than
ours does, though the overlap we measure---forgeries scoring as more ordinary than
genuine notes---is a property of the attack, not of the similarity measure.
\emph{(4) Overhead measurement.} The overhead in Section~\ref{sec:overhead} is measured
on an in-memory ledger. A deployment would add durability and concurrency control, whose
cost we do not model.
\emph{(5) Ledger isolation.} The HMAC key must remain within the trusted runtime; if
the attacker obtains it or can invoke the genuine action path, the guarantee
collapses---the same boundary \farma{} already assumes.

\section{Conclusion}
\label{sec:conclusion}
For years, agent-memory defenses judged a memory by examining the memory---its
normalcy, its signature, its wording---and an attacker who controlled the memory could
satisfy every such check. Forged reasoning and its adaptive variant are the sharpest
expression of this dead end: an optimizing attacker reduces a state-of-the-art wording
filter to zero protection, and the threat intensifies on more capable models.
Proof-of-Execution Memory escapes the pattern by refusing to inspect memory at all: it
authenticates what the agent \emph{actually did} against a tamper-evident ledger the
attacker cannot write. Across three models, three scenarios, an adversary that targets
the defense itself, and a real LangChain deployment, \poem{} drives forged-reasoning
attack success to zero while preserving legitimate operation, at microsecond cost.
The lesson generalizes beyond this attack: do not authenticate what an agent's memory
\emph{says} it did---authenticate what it \emph{actually} did.

\section*{Code Availability}
The implementation described in this paper is open source and available at
\url{https://github.com/bithabib/ai_security} under an MIT license. The repository
contains the forged-reasoning attack and its adaptive variants, our \sentinel{}
reproduction, \poem{} and the \naivepoem{} ablation, the automated adaptive attacker,
the LangChain port, and the scripts that produce every table and figure reported here,
together with the raw result files they were generated from. The core harness depends
only on the Python standard library; \texttt{matplotlib} is required to regenerate the
figures, and the LangChain port additionally requires \texttt{langchain-core} and
\texttt{langchain-openai}.

\section*{Acknowledgment}
During the preparation of this manuscript, the authors used Claude (Anthropic) to check
and improve the grammar, spelling, and sentence structure of the text. The tool was used
solely for language editing and not to generate research content, results, or analysis.
The authors reviewed and edited all AI-assisted text and take full responsibility for the
content of this article.

\bibliographystyle{IEEEtran}
\bibliography{poem}

\end{document}